# Amending the Heston Stochastic Volatility Model to Forecast Local Motor Vehicle Crash Rates: A Case Study of Washington, D.C.

**Darren Shannon (corresponding author)**
Department of Accounting and Finance
University of Limerick, Limerick, Ireland, V94 PH93
Email: darren.shannon@ul.ie
ORCID: https://orcid.org/0000-0002-7117-5049

**Grigorios Fountas**
Transport Research Institute,
Edinburgh Napier University, Edinburgh, Scotland, EH10 5DT
Email: g.fountas@napier.ac.uk
ORCID: https://orcid.org/0000-0002-2373-4221

**HIGHLIGHTS**

- We present the case of using Stochastic Volatility modelling in Transportation Safety.

- We extend the Heston model to forecast non-seasonal crash rates in Washington, D.C.

- Our model outperforms Vasicek and ARIMA-GARCH models over the forecast period.

- Highly-accurate forecasts for 2015-2019 rates demonstrate the efficacy of our model.

- Structural breaks from the series (COVID-19) suggest further improvements are required.

**ABSTRACT**

Modelling crash rates in an urban area requires a swathe of data regarding historical and prevailing traffic volumes and crash events and characteristics. Provided that the traffic volume of urban networks is largely defined by typical work and school commute patterns, crash rates can be determined with a reasonable degree of accuracy. However, this process becomes more complicated for an area that is frequently subject to peaks and troughs in traffic volume and crash events owing to exogenous events – for example, extreme weather – rather than typical commute patterns. One such area that is particularly exposed to exogenous events is Washington, D.C., which has seen a large rise in crash events between 2009 and 2020. In this study, we adopt a forecasting model that embeds heterogeneity and temporal instability in its estimates in order to improve upon forecasting models currently used in transportation and road safety research. Specifically, we introduce a stochastic volatility model that aims to capture the nuances associated with crash rates in Washington, D.C. We determine that this model can outperform conventional forecasting models, but it does not perform well in light of the unique travel patterns exhibited throughout the COVID-19 pandemic. Nevertheless, its adaptability to the idiosyncrasies of Washington, D.C. crash rates demonstrates its ability to accurately simulate localised crash rates processes, which can be further adapted in public policy contexts to form road safety targets.

# 1. INTRODUCTION

In lieu of real-time motion capturing technology, modelling crash rates[1] in an urban environment requires reasonable insight into the mobility patterns that are prevalent in an area. These insights are typically gleaned from prevailing information regarding traffic volume, the number of registered vehicles, environmental factors, and historical information on crash events. Shannon and Fountas (2021) demonstrate using an extended Heston Stochastic Volatility model how crash rates can be accurately determined from information on historical crashes and the number of registered vehicles on roadway networks. They state that their high level of accuracy would hold provided that random variations in a crash rate process are underpinned by mild fluctuations and consistent seasonal deviations. These seasonal deviations are likely driven by commute patterns, where the crash rate process is largely determined by work commutes, school runs, and vacation periods. However, seasonality cannot be assumed in areas where traffic volume (and correspondingly, traffic safety) are largely influenced by exogenous events such as adverse weather conditions or social events that attract a significant volume of traffic.

Frequent updates to the crash rate process, interspersed by disruptive events that serve to introduce multiple discontinuities, represent a distinct challenge for transportation safety researchers. In this research study, we take an interdisciplinary approach by combining the fields of quantitative finance and transportation safety to formulate a model that can provide enhanced, idiosyncratic predictions for localities that observe frequent disruptions to conventional crash rate dynamics. One such area is Washington, D.C. – a district that is comprised entirely of urban road networks, is affected by adverse weather quite frequently in the winter, and is afflicted by sporadic deviations from conventional commute patterns during national holiday periods. We accomplish our goal by forming a statistical model that embraces heterogeneous randomness in the frequency of crash events. In doing so, we at least partially offset issues relating to unobserved heterogeneity in the recorded data; a challenge that frequently affects transportation research models (Mannering *et al.* 2016).

Our approach extends the advancements made in Shannon and Fountas (2021) by introducing an amended Heston Stochastic Volatility model with pre-specified hurdle parameters to forecast crash rates. The parameters we include in the amended Heston model are a latent measure of the factors that influence crash events. In other words, we assume that the raw data we examine adequately captures the effect played by driver behaviour, vehicle and roadway characteristics, and environmental conditions.

The Heston stochastic volatility model is often used in financial settings as a means of forecasting a financial asset's plausible price evolution over a set period of time. Movements in financial asset

---

[1] The frequency of crash events after accounting for the risk exposure of vehicles in some way

prices are traditionally ascribed as being random, driven by conflicting human beliefs on the expected future value of an asset and it's relation to prevailing market information. The primary advancement attributed to the Heston model is that it combined two stochastic processes into a single model. One series produces plausible simulations for the random variability of asset prices with each time step, while the other produces plausible simulations for the extent of the dispersions associated with the asset price over a certain period of time. The combination of these two intertwined time series served to largely mimic the underlying behaviour of the variability within financial asset prices over the forecast period.

The adaption of the Heston stochastic volatility model for transportation safety is appropriate as similar dynamics have been observed in crash rates. Crash frequencies are inherently random (Lord and Mannering 2010, Mannering and Bhat 2014), as are traffic congestion patterns (Li and Chen 2017), and both are defined by conflicting human driver behaviours acting on information relating to prevailing traffic conditions and transport route options (Laval 2011).Furthermore, crashes and traffic volume are largely intertwined. Crash frequencies are highly-associated with the level of traffic congestion (acting as a proxy for the level of risk exposure associated with each roadway participant), whereas the relationship between crash rates and the level of traffic congestion is highly negative (Anastasopoulos *et al.* 2012, Guo *et al.* 2019). In other words, as the density of traffic increases, we would expect more crashes to occur in absolute terms, but we would expect fewer crashes to occur on a per-vehicle basis. As such, regardless of the crash process being forecasted, there is strong evidence to suggest that the correlation between the crashes and traffic volume is highly-significant, and this relationship should be accounted for in the modelling process. Furthermore, there is a need to account for the possibility that crash rate dynamics evolve and change over time. A significant volume of evidence has emerged to suggest that crash rate and injury severity dynamics are temporally unstable (Behnood and Mannering 2015, 2016, Mannering 2018, Hou *et al.* 2020, Tamakloe *et al.* 2020). There are numerous reasons for this phenomenon. Amongst others, successful road safety interventions (Mannering 2018) or road network alterations (Hou *et al.* 2020) may serve to change the nature of crash rates in an area, or wider macroeconomic conditions may serve to alter vehicle ownership and driver behaviour dynamics (Behnood and Mannering 2016).

We consider the ARIMA, ARIMA-GARCH, and the Vasicek model as alternatives in this study, due to their recent utility as forecasting approaches in transportation safety literature (Ramstedt 2008, Chen *et al.* 2011, Guo *et al.* 2014, Rajabzadeh *et al.* 2017). Our adaption of the Heston stochastic volatility model improves on prevailing modelling techniques for a number of reasons. ARIMA is similar to the Heston model in that the parameters are initially formed on historical observations and a mean-reversion property is embedded in the predictions. However, the ARIMA model suffers from the assumption of constant, normally-distributed variance over time, an assumption that contrasts with prior findings that the variance of crash rates differ across different time periods (Mannering 2018). Furthermore, the model parameters formed in the ARIMA model are a linear combination of prior values, limiting the flexibility of the predictions, whereas the Heston model uses the model parameters as a basis for introducing stochasticity and heteroscedasticity.

The ARIMA-GARCH improves on the ARIMA model by assuming non-constant and heteroscedastic variance in the ARIMA model errors, meaning that it can account for changing volatilities in crash rates over time. However, there is limited flexibility to the heteroscedasticity. The Heston model introduces further flexibility into the volatility predictions by including two extra parameters that allows the practitioner to embed beliefs regarding the long-run average volatility, and the speed with which the current level of volatility will revert to the long-run average. Furthermore, the ARIMA-GARCH modelling process remains hindered by the linear predictions associated with the ARIMA approach. The Vasicek model uses a different approach to the ARIMA-GARCH model to form its estimates. The Vasicek is a stochastic volatility model that was initially introduced to forecast short-term interest rate movements in financial money markets, however it has since been adapted for transportation safety (Rajabzadeh *et al.* 2017). While it avoids the linearity vulnerabilities associated with the ARIMA model, it has limited flexibility in the evolution of crash rates. Similar to the GARCH model, its primary function is to produce heteroscedastic volatility estimates that determine the randomly-evolving deviations in crash rates, but it has little utility in simulating crash rates themselves. The Heston stochastic volatility model accounts for the vulnerabilities in the aforementioned models. Its model parameters can be formed in sparse-data environments, and the formed parameters do not assume a fixed linear structure within past and future observations. Instead, the model parameters are used as a starting point for randomly-distributed and randomly-evolving simulations, which produce estimates for both the crash rate level and the crash rate volatility over time.

The model we propose operates by generating a large number of Monte Carlo simulations (we use 5,000 in this study). Using these simulations, we report 'forecasted crash rates' that are based on the median value of the simulations at each time step. Each forecasted rate is bounded by prediction intervals that indicate our degree of confidence that the realised rate will lie below the boundary provided. Improving the accuracy of forecasting models is necessitated by a continuing upward trend in crash events within Washington, D.C. (Figure 1). In response, the local government have initiated a districtwide scheme to reduce fatalities and serious injuries on its roads to zero (District Department of Transportation 2021). In addition to being a useful policy tool to track of the trend of crash events, more informed forecasts could *optimise* the allocation of traffic management resources and better direct road safety schemes. Therefore, accurate forecasting tools can improve operational efficiencies.

Our study is structured by first providing an overview of crash dynamics within Washington, D.C. from 2009-2020. We thereafter detail the formation of the amended Heston model, before demonstrating its efficacy in forecasting realised rates. Our penultimate section describes the benefits and drawbacks of our model, before we conclude.

## 2. MATERIALS AND METHODS

### 2.1. Data Description

The data used in this study is drawn from two sources – one each for reported crashes and vehicle miles travelled (acting as a proxy for traffic volume). To ensure the reported crashes are as up-to-date as possible, crash data is taken from Open Data DC (2021), an open-source government initiative that enhances data transparency and accountability in the area. Despite the valuable insights the Open Data DC dataset provides, a number of reliability issues and discrepancies persist[2]. Nevertheless, Open Data DC is used as a data source as it offers a continuous and uninterrupted stream of daily collision data from 2009-2020. As of the 1$^{st}$ January, 2021, 241,975 crash events had been recorded in Washington, D.C. since 2009. The crash events comprise property damage only, crashes involving injuries, and fatality-related crashes, and are categorised based on driver, passenger, cyclist, or pedestrian involvement.

Figure 1 demonstrates the evolution of crash frequencies in Washington, D.C. from 2009-2020. The proportion of crash severity types remains largely consistent throughout the data sample, with Property Damage Only crashes accounting for, on average, 64.3% of all crashes, injury-causing crashes accounting for 35.5% of crashes, and fatal crashes accounting for 0.2% of crashes. There is a persistent upward trend in overall crashes through the early 2010s, which plateaus in the mid-late 2010s. 2020 saw a reduction in crashes due to mobility restrictions imposed during the COVID-19 pandemic.

---

[2] There are numerous discrepancies between Open Data Dc, 2021. Crashes in dc. Washington, DC. data and alternative sources:
1. The recording of crashes in Open Data DC stretches back to 2008. However, crashes from the first half of 2008 were only partially-recorded. The unreliability of data from this period meant that 2008 data was removed from consideration, and we only consider 2009-2020 data.
2. Even after accounting for missing data in 2008, the data retrieved from Open Data DC may not contain all recorded incidents that occurred. For example, the monthly data for 2015 is consistently 10-15% lower than the officially-reported statistics submitted to the District Department of Transportation by researchers at the Howard University Transportation Research Center, 2016. Traffic safety statistics report for the district of columbia (2013-2015). Washington, DC..
3. A further discrepancy from the logged events in Open Data DC is found in fatality data. The Metropolitan Police Department, 2021. Safety & prevention | traffic fatalities. Washington, D.C. report 348 road fatalities between 2009-2020, while the logged events in the Open Data DC data specifies there were 465 fatal crashes between 2009-2020, involving 478 fatalities.

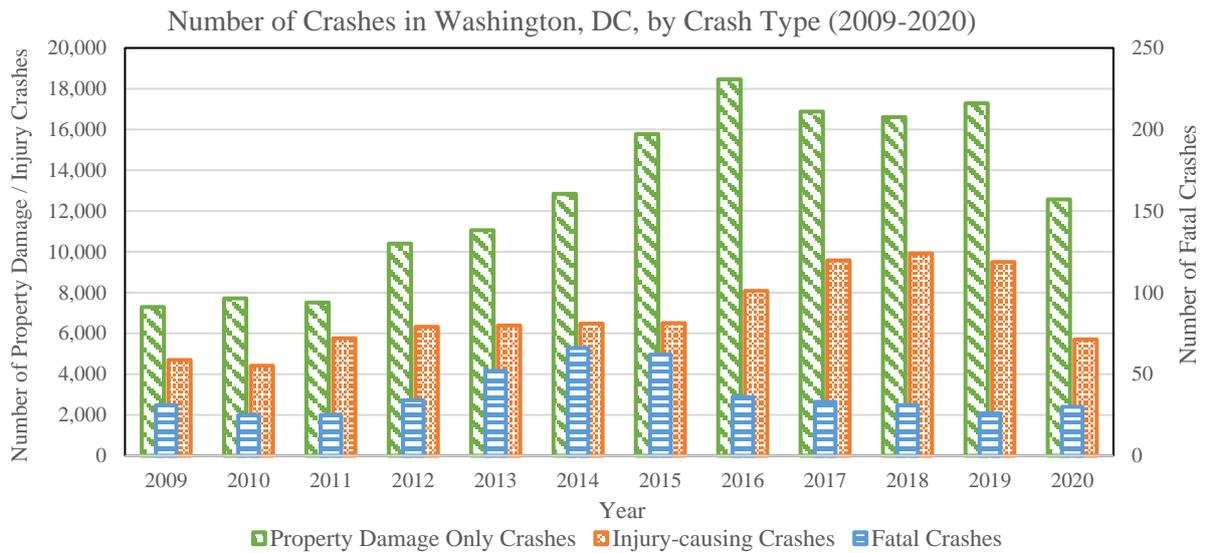

**Figure 1 Despite the upward linear trend in crash events throughout the 2010s, the breakdown of crashes by crash severity remains consistent.**

While data on crash events are extracted from Open Data DC, data on traffic volumes[3] for the period 2009-2020 are retrieved from monthly travel reports provided by the US Department of Transportation's Office of Highway Policy Information (2021). Traffic volume warrants close consideration in this study given the long-standing relationship between travel distance and crash frequency (Jovanis and Chang 1986). Travel distance reflects the length of time with which vehicle occupants are exposed to the possibility of a collision – the longer distance travelled, the greater the likelihood that a crash event may occur (Wolfe 1982). Figure 2 demonstrates the evolution of crash events in Washington, DC in conjunction with the number of vehicle miles travelled (measured in millions) on a monthly basis. Despite the variability in the crash data from 2009-2020 as highlighted in Figure 1 and Figure 2, traffic volume data maintains a stable level across the same period, albeit with persistent trough patterns in July.

---

[3] As measured by vehicle-miles travelled on Washington, DC roads

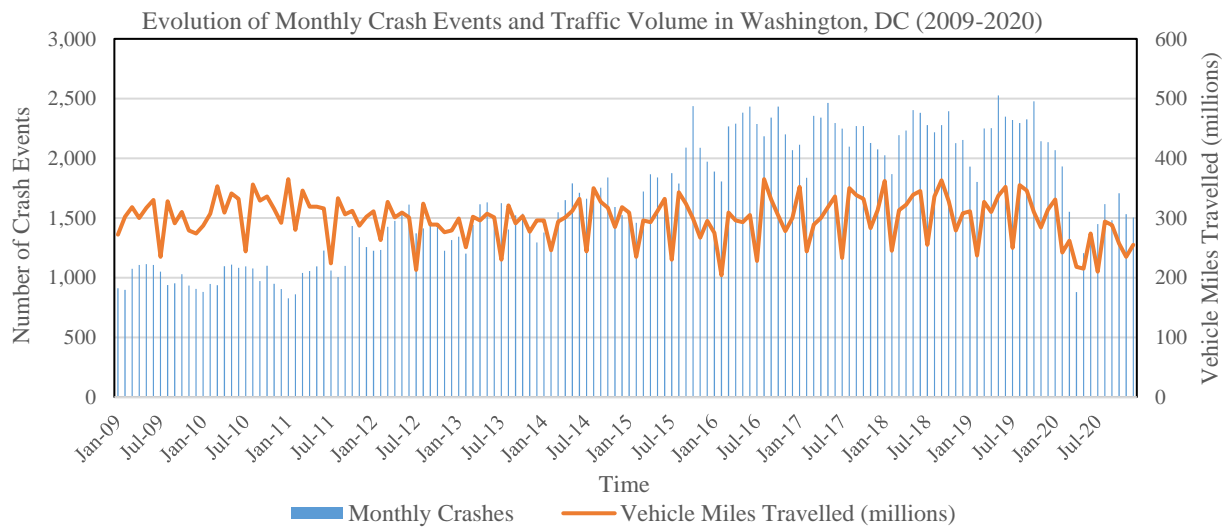

**Figure 2 The 2010s saw a doubling of crash events (bar chart), but traffic volume remained stable during the same period (line chart). The drop-off throughout 2020 was due to mobility restrictions imposed during the COVID-19 pandemic.**

To better capture the risk exposure of vehicles on transport networks, we generate a crash rate process that accounts for the dynamic relationship between traffic volume and crash event frequency. We measure the monthly 'crash rate' as the number of crash events recorded each month, taken as a function of every 1,000-vehicle miles travelled. In other words, $Crash\ Rate = \frac{Crash\ Events}{VMT\ ('000s)}$.

The resulting process is provided in Figure 3. By setting the crash rate to be our variable of interest rather than absolute number of crashes or traffic volume, we are implicitly presuming that the long-standing relationship between traffic volume and crashes holds, and can account for the idiosyncrasies associated with crash event patterns in Washington, D.C. The linear trend in the data signifies that crash rates increase by over 10% each year, on average, over the period 2009-2019. However, Figure 3 demonstrates that there are three phases within the data that are all punctuated by a form of non-stationarity. The period 2010-2014 shows a linearly increasing trend in crash rates, albeit with relatively stable variability within the data. The period 2015-2019, meanwhile, shows a relative plateau in crash rates, albeit with an increased level of variability in the data. The third process, which represents a structural break in the crash rate series, describes the effect of COVID-19 mobility restrictions on crash rates in 2020 – a linearly-decreasing trend with heterogeneous crash rate variations.

We form our analysis by forecasting crash rates over the 5-year period 2015-2019 – the second phase of crash rates in our data. By examining the second phase of data, we maintain the ability to generate crash rate forecasts over a 5-year period (2015-2019) using model parameters derived from the immediately preceding 5-year period (2010-2014), while examining the efficacy of a forecasting model prior to a significant structural break (2020). Our 'training period' therefore comprises 60 monthly crash event and traffic volume observations extending from 2010-2014 (Table 1). Table 1 reports the crash rates throughout the 60 months, along with a breakdown of how they inform the parameters of the forecasting model. The raw values are supplemented with summary statistics describing the variability of the 2010-2014 time series. The corresponding crash rate and time series characteristics for our 'testing period' (2015-2019) are available in Table A1.

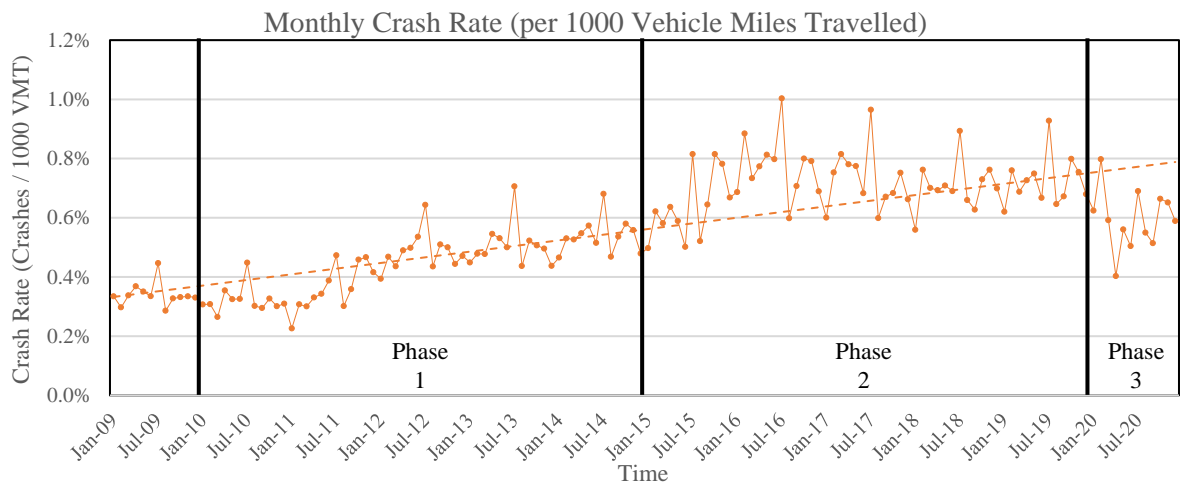

**Figure 3 The crash rate is the number of crash events that occur each month after accounting for traffic volume. The crash rate increases over time since crash events increased by 10% on average each year while traffic volumes remained relatively constant (Figure 2).**

Over the period 2010-2014, the lowest number of crashes occurred in the month of January 2011, while the highest number of recorded crashes occurred in October 2014 (Table 1). The lowest traffic volume was recorded in July 2012 with a maximum being attained in January 2011. Crash rates typically reach their lowest levels each January over the 5-year period, while reaching consistent highs each July. The lowest crash rate was recorded in January 2011, while the highest was recorded in July 2013. We note the presence of large month-on-month variations across the monthly crash rates, which we measure on a natural log-differenced scale to maintain symmetry when reporting the size of deviations relative to their starting position[4]. The largest month-on-month rise and fall in crash rates was recorded between June and August 2013: +34.45% between June and July 2013, and -47.96% between July and August 2013. The high levels of variation in the data signify the existence of high levels of volatility[5], which we report on an annualised scale $(\sigma_s\sqrt{12})$. The standard deviation values $\sigma_s$ are calculated using the values in Table 1 as $\sigma_s = \sqrt{\frac{\sum(x_i-\bar{x})^2}{n-1}}$. The 5-year volatility, after considering all rates recorded between 2010 and 2014, is 63.33%. Within-year volatilities are also calculated in order to measure the 'volatility of volatility' over the 5-year period. The 5-year volatility of volatility demonstrates the extent to which the volatility in monthly rates changes each year, and is computed as being 26.26%. The contrast between the high 5-year volatility and the lower 5-year volatility of volatility indicates that stability can be found in the instability of the process – high volatility remains a constant feature throughout the 5-year period. Both of these measures are incorporated into our forecasting model.

---

[4] For example, taking the natural log-difference of successive rates means that a 5% fall in rates, followed by a 5% rise in rates, will return the value back to its starting position.
[5] Volatility is a measure of a time series' standard deviation within a specific period of time.

**TABLE 1 Summary statistics and discerning model parameters from a time series of monthly collision data, 2010-2014.**

| | Month | Underlying Crash Rates | | | Seasonality | Volatility | | |
|---|---|---|---|---|---|---|---|---|
| | | Crashes | Vehicle Miles Travelled ('000s) | Crash Rate (Figure 2) | Deviations from Yearly Average (Figure 3) | Crash Rate Log-differences | Yearly Volatility | Yearly Volatility Log-differences |
| | Min | 827 | 213,000 | 0.227% | -37.81% | -47.96% | | |
| | Max | 1,839 | 365,000 | 0.706% | 39.17% | 34.45% | | |
| | Mean | 1,318 | 301,467 | 0.444% | 0 | 0.62% | | |
| | Std. Dev | 264 | 32,105 | 0.106% | 14.58% | 18.21% | | |
| 2010 | Jan | 881 | 287,000 | 0.307% | -4.84% | -7.43% | 65.72% | - |
| | Feb | 947 | 307,000 | 0.308% | -4.38% | 0.49% | | |
| | Mar | 936 | 353,000 | 0.265% | -17.80% | -15.13% | | |
| | Apr | 1,096 | 309,000 | 0.355% | 9.95% | 29.09% | | |
| | May | 1,109 | 341,000 | 0.325% | 0.82% | -8.67% | | |
| | June | 1,082 | 332,000 | 0.326% | 1.03% | 0.21% | | |
| | July | 1,094 | 244,000 | 0.448% | 38.99% | 31.90% | | |
| | Aug | 1,077 | 356,000 | 0.303% | -6.22% | -39.34% | | |
| | Sep | 971 | 329,000 | 0.295% | -8.51% | -2.47% | | |
| | Oct | 1,100 | 336,000 | 0.327% | 1.49% | 10.37% | | |
| | Nov | 949 | 315,000 | 0.301% | -6.61% | -8.31% | | |
| | Dec | 905 | 292,000 | 0.310% | -3.92% | 2.83% | | |
| 2011 | Jan | 827 | 365,000 | 0.227% | -37.81% | -31.33% | 77.87% | 16.97% |
| | Feb | 861 | 280,000 | 0.308% | -15.60% | 30.54% | | |
| | Mar | 1,040 | 346,000 | 0.301% | -17.50% | -2.28% | | |
| | Apr | 1,055 | 319,000 | 0.331% | -9.23% | 9.56% | | |
| | May | 1,095 | 319,000 | 0.343% | -5.79% | 3.72% | | |
| | June | 1,227 | 316,000 | 0.388% | 6.57% | 12.33% | | |
| | July | 1,060 | 224,000 | 0.473% | 29.88% | 19.78% | | |
| | Aug | 1,005 | 333,000 | 0.302% | -17.17% | -44.98% | | |
| | Sep | 1,098 | 306,000 | 0.359% | -1.52% | 17.31% | | |
| | Oct | 1,432 | 312,000 | 0.459% | 25.97% | 24.62% | | |
| | Nov | 1,339 | 287,000 | 0.467% | 28.05% | 1.64% | | |
| | Dec | 1,256 | 302,000 | 0.416% | 14.15% | -11.49% | | |
| 2012 | Jan | 1,225 | 311,000 | 0.394% | -18.89% | -5.44% | 57.72% | -33.48% |
| | Feb | 1,232 | 263,000 | 0.468% | -3.54% | 17.33% | | |
| | Mar | 1,426 | 327,000 | 0.436% | -10.21% | -7.16% | | |
| | Apr | 1,475 | 301,000 | 0.490% | 0.90% | 11.66% | | |
| | May | 1,540 | 309,000 | 0.498% | 2.62% | 1.69% | | |
| | June | 1,612 | 301,000 | 0.536% | 10.28% | 7.19% | | |
| | July | 1,371 | 213,000 | 0.644% | 32.54% | 18.39% | | |
| | Aug | 1,412 | 324,000 | 0.436% | -10.26% | -39.00% | | |
| | Sep | 1,474 | 289,000 | 0.510% | 5.02% | 15.73% | | |
| | Oct | 1,446 | 289,000 | 0.500% | 3.03% | -1.92% | | |
| | Nov | 1,226 | 276,000 | 0.444% | -8.53% | -11.90% | | |
| | Dec | 1,315 | 279,000 | 0.471% | -2.95% | 5.93% | | |
| 2013 | Jan | 1,343 | 299,000 | 0.449% | -11.47% | -4.82% | 59.83% | 20.18% |
| | Feb | 1,201 | 251,000 | 0.478% | -5.69% | 6.32% | | |
| | Mar | 1,442 | 302,000 | 0.477% | -5.89% | -0.21% | | |
| | Apr | 1,615 | 296,000 | 0.546% | 7.54% | 13.34% | | |
| | May | 1,630 | 307,000 | 0.531% | 4.65% | -2.72% | | |
| | June | 1,506 | 301,000 | 0.500% | -1.39% | -5.94% | | |
| | July | 1,624 | 230,000 | 0.706% | 39.17% | 34.45% | | |
| | Aug | 1,403 | 321,000 | 0.437% | -13.85% | -47.96% | | |
| | Sep | 1,522 | 291,000 | 0.523% | 3.09% | 17.95% | | |
| | Oct | 1,535 | 303,000 | 0.507% | -0.15% | -3.19% | | |
| | Nov | 1,374 | 277,000 | 0.496% | -2.23% | -2.11% | | |
| | Dec | 1,295 | 296,000 | 0.438% | -13.77% | -12.56% | | |
| 2014 | Jan | 1,378 | 296,000 | 0.466% | -13.56% | 6.21% | 57.35% | -17.29% |
| | Feb | 1,305 | 246,000 | 0.530% | -1.50% | 13.06% | | |
| | Mar | 1,548 | 294,000 | 0.527% | -2.23% | -0.75% | | |
| | Apr | 1,649 | 301,000 | 0.548% | 1.73% | 3.97% | | |
| | May | 1,789 | 312,000 | 0.573% | 6.47% | 4.56% | | |
| | June | 1,711 | 332,000 | 0.515% | -4.30% | -10.67% | | |
| | July | 1,661 | 244,000 | 0.681% | 26.40% | 27.83% | | |
| | Aug | 1,640 | 350,000 | 0.469% | -12.99% | -37.35% | | |
| | Sep | 1,753 | 327,000 | 0.536% | -0.46% | 13.46% | | |
| | Oct | 1,839 | 317,000 | 0.580% | 7.72% | 7.90% | | |
| | Nov | 1,592 | 285,000 | 0.559% | 3.72% | -3.78% | | |
| | Dec | 1,524 | 318,000 | 0.479% | -11.01% | -15.32% | | |
| Model Inputs | | January 2015 Rate 0.498% | | 5-year Annual Growth Rate 13.61% | | | 5-year Volatility 63.33% | 5-year Volatility of Volatility 26.26% |

In order to detect seasonality in the crash rate data, Figure 4 captures the deviation of each month's crash rate from their corresponding yearly average. These deviations are categorised by month in the form of a clustered bar chart across the 2009-2020 period. We note little evidence of a seasonal pattern in our time series data. However, there are persistent deviations in crash rate associated with certain months, with periodic spikes visible each January (trough), July (peak), and August (trough) (Figure 4).

January typically sees a fall in crash events despite a high level of traffic volume, resulting in a 20% drop in crash rates relative to the annual average. It is possible that the adverse weather conditions typically seen in Washington, D.C. during this period, such as snow and ice, leads to an increased level of risk compensation amongst drivers. Prior research has demonstrated that drivers alter their driving habits in the face of adverse weather conditions (Saha *et al.* 2016), and that snow-covered roads produce the highest level of perceived risk amongst drivers (Hjelkrem and Ryeng 2016). Although research has pointed out that lower-severity crash rates tends to increase during snowy periods (Strong *et al.* 2010), it has also been found that the effect is more pronounced in rural areas than urban areas (Andrey *et al.* 2013). Given that the Washington, D.C. road networks are comprised entirely of urban roads, the increased level of risk compensation may outweigh the increased level of crash risk. The increased level of caution exhibited while driving mean that while traffic volumes remain in line with the surrounding months, the number of crash events fall, resulting in a downward spike in the number of crashes per 1,000 VMT each January.

Conversely, a reduced level of risk compensation may be an explanatory factor in the upward spike seen in July crash rates (Figure 4). This month typically sees a large reduction in traffic volume, yet the number of crash events remain relatively constant (Figure 2). The net effect is a significant increase (+30% relative to the annual average) in the crash rate. The rise in the crash rate for July can be at least partially attributed to the 4$^{th}$ of July celebrations, which sees a significant increase in reckless driving in spite of the lower traffic volume. Across the United States, alcohol-impaired driving is higher than the annual average over the Independence Day celebration period, and fatality rates spike relative to equivalent non-holiday periods (National Safety Council 2021). Washington, D.C. sees a similar pattern in its crash statistics (Government of the District of Columbia 2021). Immediately following the July holiday period, August sees as a significant increase in traffic volume relative to the surrounding months, although crash events typically remain level with July and September. A constant number of crash events relative to higher traffic volumes means that the crash rate falls by 13% relative to the annual average.

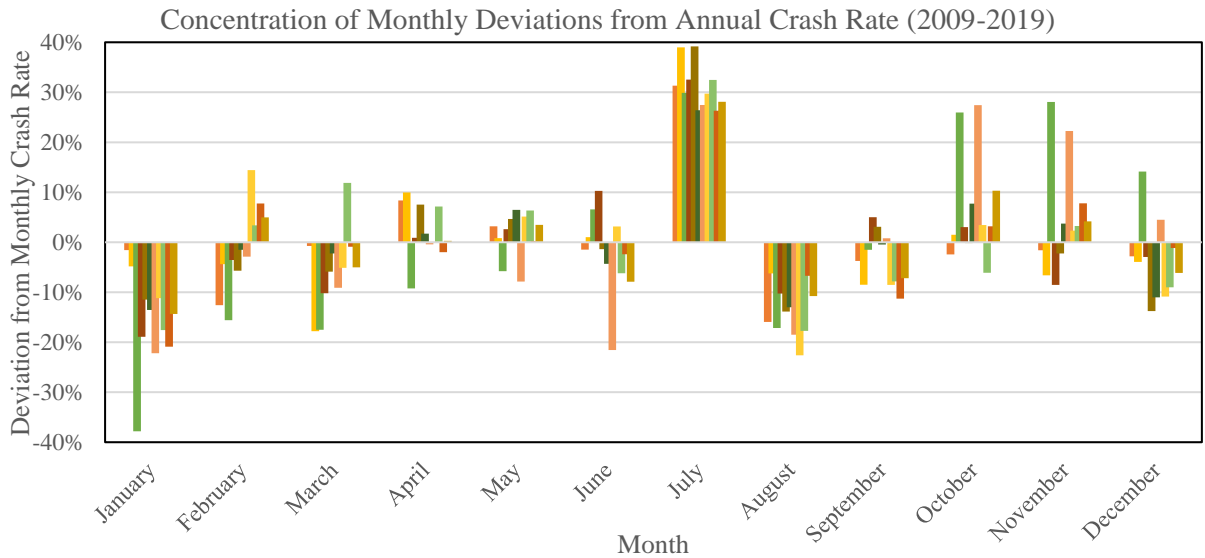

**Figure 4 Clustered bar charts of monthly crash rates from 2009-2019, represented by their deviation from their respective yearly average. January (-20%), July (+30%), and August (-13%) see consistent deviations in crash rates relative to their yearly average.**

### 2.2. Methodology: the Amended Heston Model

In this study, we make amendments to the Heston stochastic volatility model (Heston 1993) to make it suitable to model monthly crash rates. An examination of the parameters in the original Heston model are available in Shannon and Fountas (2021), who tackle a similar issue but propose a different approach in order to capture seasonal crash characteristics. The Heston stochastic volatility model is effective at capturing random processes whose absolute values follow a lognormal distribution, and whose successive natural-log differences follow a normal distribution. These traits are found in Washington, D.C. crash rates (Figure A1) and agrees with prior examinations of crash rate distributions (Ma *et al.* 2016).

The Heston model also allows for the correlation between crash rates and the volatility of crash rates to be incorporated in to the model estimates – a favourable trait given the long-standing relationship between crash event frequency and traffic volume. Furthermore, it does not require the standard deviation $\left(\sqrt{v_t}\right)$ of the underlying process to remain constant over time. Instead, the standard deviation is assumed to be a stochastic process itself, which reverts to an estimated average over time. This latter characteristic of the model agrees with the temporal instability assumption among crash rates (Mannering 2018), wherein different time periods exhibit different levels of fluctuations in crash rates.

The amended Heston model, for the purposes of this study, calculates stepwise changes in the underlying collision rates $C_t$ by the equation:

$$dC_t = \left(\mu C_1 dt + \sqrt{v_t} C_1 dW_t^C\right) + \left(\overline{C_t^Y} \times G_t\right) \quad (1)$$

Where $v_t$, the instantaneous variance or squared volatility, is a mean-reverting stochastic process whose mean reversion rate is determined by the practitioner. Each change in $v_t$ is represented as

$$dv_t = \kappa(\theta - v_t)dt + \xi\sqrt{v_t}dW_t^v \quad (2)$$

The $\overline{C_t^Y} \times G_t$ term is an adjustment that is added to account for the periodic spikes in crash rates – an amendment to the model based on the findings in Figure 4. Both $W_t^C$ and $W_t^v$ signify Wiener processes with correlation $\rho$ – that is, stochastic processes with independent and identically-distributed observations drawn from a normal distribution with mean 0. The level of correlation is found by computing the level of association between annual crash rates and annual crash rate volatility. The remaining parameters can be reasoned as follows:

- $\mu$ is set to equal the expected average annual growth rate in crash rates, where expectations on growth are discerned using historical crash rates.
- $\theta$ is the long-run variance, or the rate to which month-to-month fluctuations in crash rates will tend to over time. As $t$ tends to infinity, the month-to-month variances $v_t$ are expected to revert to $\theta$.
- $\kappa$ is the mean-reversion rate, or the rate at which the prevailing variance of crash rates $v_t$ reverts to $\theta$.
- $\xi$ is the volatility of volatility, which determines the extent of the fluctuations in each successive $v_t$. The 'volatility of volatility' parameter allows practitioners to control the level of heterogeneity within the model – one period may see small random fluctuations in monthly predictions, while the next period may see large random fluctuations in monthly predictions. This accords with prior research that has identified instabilities in crash rate processes over time due to fundamental shifts in underlying travel behaviours (Behnood and Mannering 2016). For example, economic downturns may affect vehicle ownership and employment levels which would, in turn, affect travel behaviours. As such, we embed the assumption that the extent of the fluctuations in month-to-month crash rates ($\sqrt{v_t}$) vary randomly over time, in addition to the assumption that the fluctuations vary randomly over time.

### 2.2.1. Amendments: Crash and Variance Steps, and Spike Adjustments

Two core amendments are made to the original Heston (1993) model in this study. In line with (Shannon and Fountas 2021), the first deviation is a departure from state-dependent changes in stepwise crash rates $C_t$. The original Heston model assumed that asset values followed an Itô Process, wherein stepwise increments ($C_{t+1}$) are scaled to the size of the preceding value ($C_t$). In this way, if asset values were to reach near-zero values, the state-dependent process would ensure a high-likelihood of the values being 'trapped' near-zero, since a near-zero change will occur at the following time step. This makes sense for asset values that may cease to exist once their value reaches zero (e.g., in the event of bankruptcy). However, this is counter-intuitive in the context of monthly crash rates, which are largely dependent on travel patterns and traffic volumes (Wolfe 1982, Jovanis and Chang 1986, Commandeur *et al.* 2013, Regev *et al.* 2018), but are otherwise unrelated enough to be assumed as temporally independent. As such, we scale successive values in the series only to the initial value $C_1$. We further impose a condition that in the unlikely event that a negative crash rate is predicted when a simulation is near-zero, the absolute value of the prediction is used (a 'reflection' scheme), rather than setting these values to zero (a 'truncation' scheme). Furthermore, while we assume that crash rates are state-independent, we assume that stepwise changes in variances ($v_t$) are state-dependent. We retain this assumption on the basis that the variance in monthly crash rates will trend toward a long-term average $\theta$ over time. As such, we retain the conjecture that our uncertainty as to the size of future variance movements is proportional to the level of the prevailing variance. Finally, we impose the constraint $\kappa > \frac{\xi^2}{2\theta}$ (the Feller Condition) that prevents negative volatilities from being generated by the model (Grzelak and Oosterlee 2011).

The second amendment is the addition of an adjustment to the prevailing-year estimates to account for expected intermittent spikes in crash rates. Our amendment is similar to that of a hurdle model, which specifies one process when 'zero' is expected, and another process expected values that are greater than zero. Hurdle models have previously been employed in transportation safety literature (Ma *et al.* 2016). In our case, however, the hurdle does not abide by a probability density function. Instead, we 'hardcode' the probability density function to activate for months January, July, and August. This is to take into account the consistent wavelet patterns that form around these months during the period 2009-2020 (Figure 3).

In **Equation 1**, $\overline{C_t^Y}$ denotes the prevailing calendar-year average (January-December) for simulated monthly crash rates. $G_t$ incorporates the 'spike' adjustment based on the placement of $C_t$ within the calendar year, which is represented by:

$$G_t = \begin{cases} \mod(t,12) = 1 \Rightarrow \alpha \\ \mod(t,12) = 7 \Rightarrow \beta \\ \mod(t,12) = 8 \Rightarrow \gamma \\ \alpha, \beta, \gamma \in N(\mathbb{R}[a,b]) \\ \mod(t,12) \notin [1,7,8] \Rightarrow 0 \end{cases} \quad (3)$$

In this case, the spike adjustment $G_t$ determines whether the crash rates are adjusted by a rate of α, β, or γ, or remains at the standard rate. Once the 'hurdle' is crossed, the random innovations (i.e., the adjustment parameters α, β, and γ) are normally-distributed real numbers drawn from $N(\mathbb{R}[a,b])$. $a$ and $b$ are unique to each month, and correspond to the mean and standard deviation of departures from the annual average rate as recorded over the preceding 5-year period, respectively. For example, over the period 2010-2014, January crash rates in Washington, D.C. fell on average 17.3% relative to annual averages, with a standard deviation of 12.5%. July rates, on the other hand, increased 33.4% on average over the 5 years, with a standard deviation of 5.6%.

## 3. RESULTS

### 3.1. The Case for 2015-2019 Results

The results of the amended Heston model used to forecast collision rates for 2015-2019 are provided in Figure 5. All calculations are formed in Matlab. The scenario presented in Figure 5 is based on the median values of 5,000 random simulations. As derived in Table 1, the annual drift term for 2015-2019 is based on the average annual growth in crash rates between 2010-2014 ($\mu = 13.61\%$). We assume there will be no change in this volatility measure over the 5-year period, and hence set the long-run average volatility to be equal to the instantaneous volatility (i.e. $v_0 = \theta = 63.33\%$). As mentioned previously, there is stability to the instability of crash rates – while the volatilities are high each year, the year-to-year changes in volatility are relatively low. The volatility of volatility measure $\xi$ is calculated as 26.26% based on fluctuations in volatility throughout the 2010-2014 period. To satisfy the Feller Condition and prevent negative variances in the simulations, the mean reversion rate $\kappa$ is set to 5.45%[6]. We also note the strong negative correlation between crash rates and traffic volume for the period 2010-2014; we set $\rho = -59.36\%$. This will result in simulations of a stochastic process where higher volatilities are associated with lower crash rates, and vice versa, and accords with prior literature that found an inverse relationship between traffic volume and crash rates (Anastasopoulos *et al.* 2012, Zeng *et al.* 2018, Guo *et al.* 2019). Parameters α, β, and γ are set such that they have mean values -17.3%, +33.4%, and -12.1%, and standard deviation values 12.5%, 5.6%, and 4.1%, respectively.

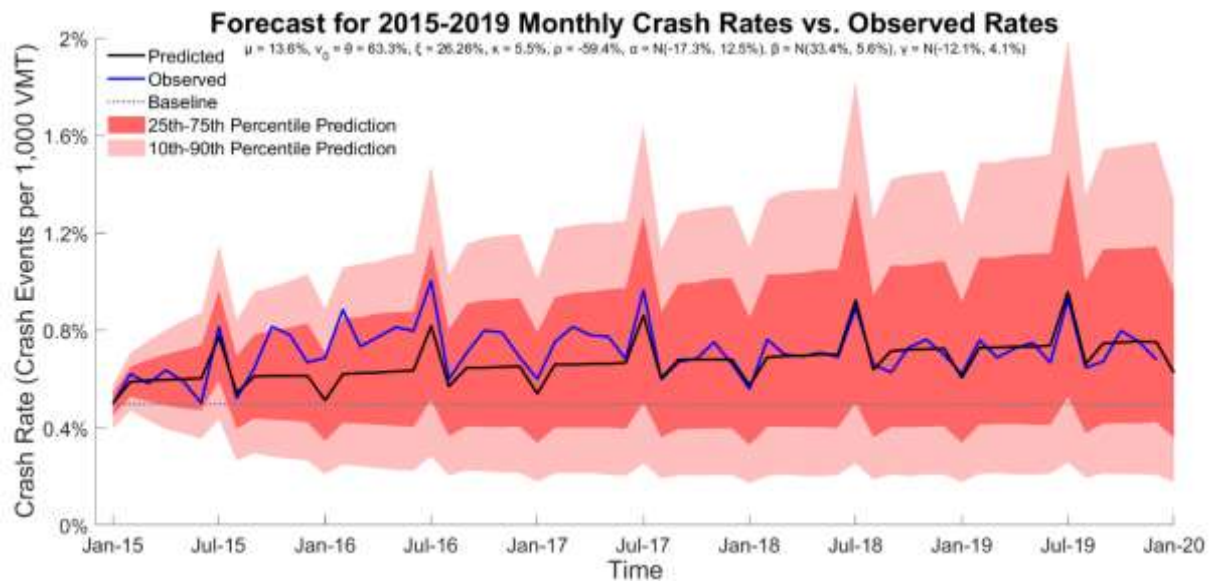

**Figure 5 Predicted vs. Observed monthly crash rates from 2015-2019. Predictions are formed using the amended Heston model, using parameters from crash events between 2010-2014.**

---

[6] The mean-reversion measure is redundant here given that the instantaneous variance = the long-run variance.

The baseline rate indicates that 0.498% of vehicles are involved in a crash for every 1,000 vehicle-miles travelled. As detailed in Figure 2, Washington, D.C. saw a consistent rise in crash frequency over the study period, despite traffic volume remaining periodically constant. Using the above parameters, the estimates from the amended Heston model align closely with observed rates for the following five years (Figure 5). The median simulation predicts an upward trend from 2015 in line with realised rates, and the embedded periodic spikes in the simulations track closely to observed spikes. A significant deviation between simulated and observed rates was observed between autumn 2015 to summer 2016, however, which can partially be explained by adverse weather events that affected Washington, D.C. throughout early 2016 and enhanced data collection efforts that began in 2015 (Government of the District of Columbia 2015). While the model simulated an increase in crash rates, during this period, there was a larger than expected spike in recorded crash frequency coinciding with an atypical drop-off in traffic volume.

Nevertheless, over the course of the 5-year forecast period, only two months saw crash rates breach the 50% prediction interval. Error statistics measuring the absolute and relative (%) difference between the forecasted rates and the observed rates are available in Table 2. The results indicate a consistent year-on-year forecasting accuracy. Absolute inaccuracies remain less than 0.1%, while relative inaccuracies remain relatively consistent at approximately 4-9% each year after excluding the large deviation from expected rates between late 2015 and summer 2016. The 5-year monthly average forecasting accuracy of 91.4% highlights the efficacy of the amended Heston model in forecasting monthly crash rates in a locality typified by frequent deviations from typical commute and crash patterns. However, it must be noted that, on the basis of the 2015 and 2016 results, the model struggles when both crash frequencies and traffic volumes significantly diverge.

**TABLE 2 Error statistics for Heston forecasts of monthly collision rates between 2015 and 2019 (5000 simulations).**

|  | Mean Absolute Error (MAE) | Root Mean Squared Error (RMSE) | Mean Absolute Percentage Error (MAPE) |
| --- | --- | --- | --- |
| 2015 | 0.060% | 0.086% | 8.77% |
| 2016 | 0.137% | 0.152% | 17.14% |
| 2017 | 0.063% | 0.081% | 8.21% |
| 2018 | 0.027% | 0.038% | 3.91% |
| 2019 | 0.035% | 0.043% | 4.98% |
| Average | 0.065% | 0.090% | 8.60% |

### 3.2. Amended Heston vs. Alternative Forecasting Methods

Forecasting methods have long-been employed in road safety and traffic flow dynamics literature, with many approaches detailed in Washington *et al.* (2020). Variants of ARIMA models (Ramstedt 2008), GARCH models (Zhang *et al.* 2013), or combinations of both (Guo *et al.* 2014) are common. Forecasts on the stochastic nature of traffic flows were also achieved through the use of the Vasicek model, a stochastic process adopted from quantitative finance (Rajabzadeh *et al.* 2017). To ensure comparability between the amended Heston model and the Vasicek model, the same amendments we propose are applied to the methodology used in Rajabzadeh *et al.* (2017), such than an 'amended Vasicek model'[7] is created. ARIMA and ARIMA-GARCH models are also applied to measure their effectiveness at forecasting crash rates. Full details on the performance of the amended Vasicek model, the ARIMA model, and the ARIMA-GARCH models are available in Table 3 and Figure 6. The results demonstrate that the amended Heston model provides a superior performance than the other models in out-of-sample forecasting accuracy tests, producing the closest affinity to 2015-2019 collision rates (Table 3, Figure 6), based on 2010-2014 parameters.

**TABLE 3 Error statistics describing the relative (%) differences between forecasted and observed monthly crash rates between 2015 and 2019. The amended Heston model proved the most consistent year-on-year.**

|  | Mean Absolute Percentage Error (MAPE) | | | |
| --- | --- | --- | --- | --- |
|  | Amended Heston *(from Table 2)* | Amended Vasicek | ARIMA *Optimal fit:* (1,2,2) | ARIMA-GARCH *Optimal fit:* (1,2,2) × (2,1) |
| 2015 | 8.77% | 8.98% | 10.58% | 10.58% |
| 2016 | 17.14% | 18.66% | 16.41% | 16.42% |
| 2017 | 8.21% | 9.74% | 7.04% | 7.03% |
| 2018 | 3.91% | 4.17% | 15.70% | 15.66% |
| 2019 | 4.98% | 4.84% | 27.88% | 27.82% |
| Average | 8.60% | 9.28% | 15.52% | 15.50% |

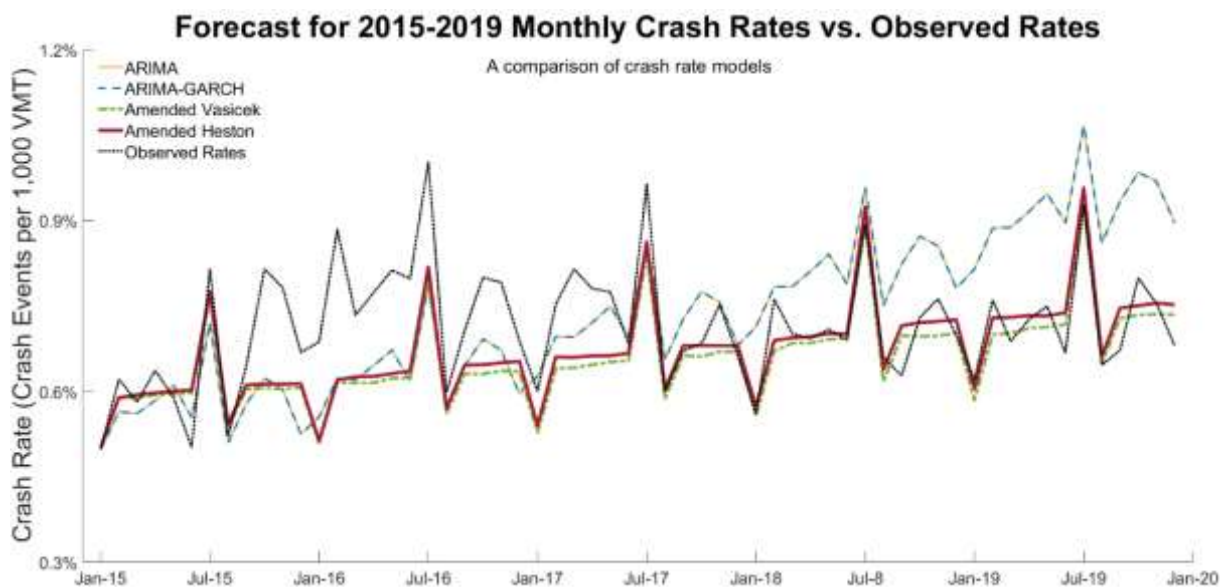

**Figure 6 Comparison of the predictions generated by all models in Table 3 for monthly crash rates from 2015-2019.**

---

[7] The adjustments include the periodic spikes, and the state-independency for crash rates.

# 4. DISCUSSION: BENEFITS AND DRAWBACKS OF STOCHASTIC VOLATILITY FOR FORECASTING

## 4.1. Benefits: Flexibility

Stochastic Volatility models are a non-deterministic means of forecasting uncertainty. They are not causal models. The amended Heston model does not base its parameter effect sizes on specific variables. Instead, it is a latent process. It assumes that the effects of the underlying variables have been captured in the raw data. However, rather than being viewed as a weakness, stochastic volatility models can be used in transportation safety contexts with great effect. Given the dynamic nature of crash rate processes, a broad indicator is more likely to be robust to temporal instabilities that induce frequent deviations from the crash rate process than a collection of indicators that form predictions on the basis of fixed effect sizes and limited amounts of in-built variability (Hou *et al.* 2020).

Our focus is on the monthly 'crash rate', which is a measure of crash events after accounting for traffic volumes. Although we demonstrate the efficacy of the amended Heston model in producing short-term forecasts of crash rates, the predominant value associated with our approach is the flexibility of the parameters in the model. We note that the model parameters used in this study are based solely on historical data. However, these parameters may be adjusted based on expert insight to account for changing patterns in crash events and traffic volumes. For example, the analysis in §2.2.1 identified January, July, and July as having consistent year-on-year spikes that were factored into the modelling process. These spikes are crash rate idiosyncrasies associated with Washington, D.C. that may not otherwise apply to different localities. However, the idiosyncratic crash characteristics associated with other localities can be similarly embedded in the model estimates, especially when there is little apparent evidence of seasonality as assumed in Shannon and Fountas (2021). In addition, based on a lack of evidence stating otherwise, we assumed the long-term variance in crash rates ($\theta$) would equal the instantaneous variance ($v_0$). It is possible to factor in beliefs regarding the long-term variance of collision rates, which have been shown to vary in different time periods (Mannering 2018). A similar rationale can be provided for the remaining parameters – for example, the spike in the volatility of crash rates with the onset of COVID-19 (Figure 3) may revert to a more stabilised long-run average.

As well as adding to the existing toolset of transportation safety modelling practitioners, the stochastic volatility approach has distinct ramifications for public policy. Granular forecasts can be determined on a local level, with limited data requirements, and while embedding temporal instability in to the estimates, to form the basis of practical road safety targets. Well-reasoned estimates, therefore, can be produced when assessing the expected impact of proposed road safety interventions or road network alterations. We preface the above by emphasising that no forecasting model is without faults. However, the advancement proposed in this study produces highly accurate forecasts with the only requirement being an insubstantial set of historical data.

### 4.2. Drawbacks: Performance during COVID-19 (2020)

We also assess the performance of forecasting models on crash rates recorded during 2020. The parameters for the amended Heston and Vasicek models are derived from observed crash rates during the period 2015-2019. The results are highlighted in Table 4, which showcases a demonstrably worse set of results for each of the forecasting models.

**TABLE 4 Error statistics for the model forecasts describing the relative (%) differences between forecasted and observed monthly collision rates for 2020. All models performed poorly with the onset of COVID-19.**

|  | Mean Absolute Percentage Error (MAPE) | | | |
|---|---|---|---|---|
|  | Amended Heston | Amended Vasicek | ARIMA *Optimal fit:* (3,2,2) | ARIMA-GARCH *Optimal fit:* (3,2,2) × (5,4) |
| 2020 | 28.95% | 27.80% | 28.86% | 25.26% |

A number of factors can explain the drop-off in model performance. 2020 saw an exogenous event (the onset of COVID-19) significantly disrupt conventional commute and travel patterns in Washington, D.C. Partial disruptions were imposed on March 13$^{th}$ 2020 with the closure of schools and a public order of dissuasion on unnecessary travel (Government of the District of Columbia 2020a). This was swiftly followed by a district-wide stay-at-home order, put in place between March 31$^{st}$ and May 29$^{th}$ 2020 (Government of the District of Columbia 2020c, b), which represented a full disruption to crash event and travel patterns. As a result, vehicle miles travelled (VMT) were consistently 20% lower in year-on-year terms from March 2020 until December 2020.

In addition to disruptions in travel patterns, there is evidence that the onset of COVID-19 negatively influenced the dynamics of driver behaviour along transport routes. Nationwide statistics indicated that fatality rates increased in comparison to 2019 rates, despite the significant fall in vehicle miles travelled. The United States recorded 1.45 fatalities per 100 million VMT in the 2$^{nd}$ quarter of 2020, a 34% year-on-year increase. Likewise, the 3$^{rd}$ quarter of 2020 saw 1.48 fatalities per 100 million VMT, a 26.5% year-on-year increase (NHTSA 2020a). It is unclear, at this juncture, the specific reasons for the rise in fatality rates. However, preliminary results noted significant deviations in the causes of crashes during 2020, relative to 2019. Crashes on urban interstates, which represent a sizable proportion of Washington, D.C. roads, increased by 15% nationwide in 2020. Night-time and weekend crashes also saw significant increases – 11% and 9%, respectively. In addition, speeding-related crashes increased by 11%, while alcohol-related crashes increased by 9% (NHTSA 2020b). The change in the manner of collisions during 2020 may point toward a greater level of reckless and aggressive driving that is not typical of commute travel patterns and may be fostered by the significantly lower traffic volumes in urban networks during the outbreak periods of the pandemic[8].

---

[8] An alternative explanation for the higher fatality rate per VMT would point to the theory of the self-selected sample Mannering, F., Bhat, C.R., Shankar, V., Abdel-Aty, M., 2020. Big data, traditional data and the tradeoffs between prediction and causality in highway-safety analysis. Analytic Methods in Accident Research 25, 100113.. According to this theory, the crash statistics captured the proportional risk associated with high-risk drivers, since it is likely that a high proportion of risk-

Nevertheless, although the Heston model demonstrates superior performance over the other conventionally-used models considered in this study, the drop in forecasting accuracy from Table 3 to Table 4 indicates that more research is required on 'black swan' events in crash rates, given the heavy reliance on using historical data to forecast future periods. Future iterations of forecasting models should focus on the affect that deviations from typical driving behaviour may have on crash rates, rather than solely basing their estimates on assuming non-deterministic and heterogeneous crash processes, while assuming stable human behaviour patterns.

## 5. CONCLUSION

Washington, D.C. presents a unique use case of a locality where roadway travel entirely subsists on urban roadways, and where crash rates are frequently affected by sporadic deviations from the norm. Its exposure to exogenous events creates a localised crash rate process that deviates from a heteroscedastic, yet consistent process of seasonality shaped by work and school commutes. Nevertheless, we have demonstrated throughout this manuscript the utility of using stochastic volatility to form accurate forecasts that improve upon conventional forecasting models. We demonstrate the efficacy of the amended Heston model in forecasting crash rates, which maintains over 90% accuracy over a 5-year period despite the periodic deviations from a stable crash rate process. However, the results in Table 4 demonstrate flaws that exist within forecasting models, an affliction that extends to the stochastic volatility model proposed in this study. A sudden, sharp disruption in collision and travel volume statistics followed the imposition of mobility restrictions in Washington, D.C. in early-mid 2020. As such, we caution against an over-reliance on its results when a structural break is detected or anticipated. Nevertheless, our results demonstrate the utility of the Heston stochastic volatility model in providing reasonable estimations for the evolution of crash rates, even in localities affected by frequent disruptions to the crash rate process. Our research, therefore, can have a desirable impact in enhancing transparency in public policy settings, since the expected outcomes of road safety interventions and roadway network alterations can be underpinned by well-informed, reasonable, and practical simulations.

### 5.1. Author Contributions

DS planned and led the drafting of the manuscript in conjunction with GF. DS coded the model, gathered the data and undertook the analysis. GF contributed to the design of the study. Both authors were involved in the interpretations of the findings and the drafting of the manuscript.

### 5.2. Declaration of Competing Interest

The authors declare that they have no known competing financial interests or personal relationships that could have appeared to influence the work reported in this paper.

---

averse drivers avoided driving where possible in 2020 (who would otherwise 'average out' the crash rate associated with the high-risk group). There is evidence against this theory, however. The number of fatalities also increased in absolute terms in Q3 2020 (by 13%), not just in relative (per VMT) terms.

# 7. APPENDIX

## 7.1. Crash and Traffic Volume Statistics: 2015-2019

**TABLE A1 Summary statistics and discerning model parameters from a time series of monthly collision data, 2015-2019.**

|  | | Underlying Crash Rates | | | Seasonality | Volatility | | |
|---|---|---|---|---|---|---|---|---|
|  | Month | Crashes | Vehicle Miles Travelled ('000s) | Crash Rate (Figure 3) | Deviations from Yearly Average (Figure 4) | Crash Rate Log-differences | Yearly Volatility | Yearly Volatility Log-differences |
| Min |  | 1,460 | 204,000 | 0.498% | -22.62% | -51.66% | | |
| Max |  | 2,526 | 365,000 | 1.003% | 32.48% | 48.52% | | |
| Mean |  | 2,146 | 303,983 | 0.715% | 0 | 0.58% | | |
| Std. Dev |  | 243 | 38,672 | 0.104% | 13.35% | 19.81% | | |
| 2015 | Jan | 1,538 | 309,000 | 0.498% | -22.18% | 3.79% | 84.06% | - |
| | Feb | 1,460 | 235,000 | 0.621% | -2.87% | 22.17% | | |
| | Mar | 1,721 | 296,000 | 0.581% | -9.10% | -6.63% | | |
| | Apr | 1,866 | 293,000 | 0.637% | -0.43% | 9.11% | | |
| | May | 1,839 | 312,000 | 0.589% | -7.85% | -7.74% | | |
| | June | 1,666 | 332,000 | 0.502% | -21.54% | -16.09% | | |
| | July | 1,875 | 230,000 | 0.815% | 27.46% | 48.52% | | |
| | Aug | 1,788 | 343,000 | 0.521% | -18.50% | -44.72% | | |
| | Sep | 2,089 | 324,000 | 0.645% | 0.80% | 21.26% | | |
| | Oct | 2,437 | 299,000 | 0.815% | 27.43% | 23.44% | | |
| | Nov | 2,088 | 267,000 | 0.782% | 22.27% | -4.14% | | |
| | Dec | 1,972 | 295,000 | 0.668% | 4.51% | -15.69% | | |
| 2016 | Jan | 1,889 | 275,000 | 0.687% | -11.17% | 2.72% | 72.90% | -14.25% |
| | Feb | 1,805 | 204,000 | 0.885% | 14.43% | 25.32% | | |
| | Mar | 2,267 | 309,000 | 0.734% | -5.12% | -18.73% | | |
| | Apr | 2,290 | 296,000 | 0.774% | 0.05% | 5.31% | | |
| | May | 2,382 | 293,000 | 0.813% | 5.14% | 4.96% | | |
| | June | 2,433 | 305,000 | 0.798% | 3.16% | -1.90% | | |
| | July | 2,287 | 228,000 | 1.003% | 29.72% | 22.91% | | |
| | Aug | 2,184 | 365,000 | 0.598% | -22.62% | -51.66% | | |
| | Sep | 2,341 | 331,000 | 0.707% | -8.54% | 16.72% | | |
| | Oct | 2,432 | 304,000 | 0.800% | 3.46% | 12.32% | | |
| | Nov | 2,200 | 278,000 | 0.791% | 2.34% | -1.09% | | |
| | Dec | 2,068 | 300,000 | 0.689% | -10.85% | -13.80% | | |
| 2017 | Jan | 2,113 | 352,000 | 0.600% | -17.58% | -13.83% | 72.28% | -0.86% |
| | Feb | 1,837 | 244,000 | 0.753% | 3.37% | 22.65% | | |
| | Mar | 2,355 | 289,000 | 0.815% | 11.89% | 7.91% | | |
| | Apr | 2,341 | 300,000 | 0.780% | 7.15% | -4.33% | | |
| | May | 2,463 | 318,000 | 0.775% | 6.35% | -0.75% | | |
| | June | 2,295 | 336,000 | 0.683% | -6.21% | -12.57% | | |
| | July | 2,248 | 233,000 | 0.965% | 32.48% | 34.54% | | |
| | Aug | 2,097 | 350,000 | 0.599% | -17.73% | -47.64% | | |
| | Sep | 2,269 | 338,000 | 0.671% | -7.82% | 11.37% | | |
| | Oct | 2,270 | 332,000 | 0.684% | -6.12% | 1.84% | | |
| | Nov | 2,128 | 283,000 | 0.752% | 3.25% | 9.51% | | |
| | Dec | 2,074 | 313,000 | 0.663% | -9.02% | -12.65% | | |
| 2018 | Jan | 2,025 | 362,000 | 0.559% | -20.90% | -16.94% | 59.59% | -19.31% |
| | Feb | 1,867 | 245,000 | 0.762% | 7.75% | 30.91% | | |
| | Mar | 2,194 | 313,000 | 0.701% | -0.88% | -8.36% | | |
| | Apr | 2,232 | 322,000 | 0.693% | -1.99% | -1.12% | | |
| | May | 2,403 | 339,000 | 0.709% | 0.23% | 2.24% | | |
| | June | 2,381 | 345,000 | 0.690% | -2.41% | -2.67% | | |
| | July | 2,278 | 255,000 | 0.893% | 26.32% | 25.81% | | |
| | Aug | 2,217 | 336,000 | 0.660% | -6.70% | -30.30% | | |
| | Sep | 2,278 | 363,000 | 0.628% | -11.26% | -5.01% | | |
| | Oct | 2,394 | 328,000 | 0.730% | 3.21% | 15.11% | | |
| | Nov | 2,127 | 279,000 | 0.762% | 7.80% | 4.35% | | |
| | Dec | 2,153 | 308,000 | 0.699% | -1.16% | -8.67% | | |
| 2019 | Jan | 1,930 | 311,000 | 0.621% | -14.33% | -11.90% | 63.33% | 6.09% |
| | Feb | 1,802 | 237,000 | 0.760% | 4.96% | 20.31% | | |
| | Mar | 2,250 | 327,000 | 0.688% | -5.01% | -9.99% | | |
| | Apr | 2,252 | 310,000 | 0.726% | 0.29% | 5.43% | | |
| | May | 2,526 | 337,000 | 0.750% | 3.47% | 3.13% | | |
| | June | 2,349 | 352,000 | 0.667% | -7.88% | -11.62% | | |
| | July | 2,320 | 250,000 | 0.928% | 28.11% | 32.97% | | |
| | Aug | 2,295 | 355,000 | 0.646% | -10.75% | -36.15% | | |
| | Sep | 2,326 | 346,000 | 0.672% | -7.20% | 3.91% | | |
| | Oct | 2,477 | 310,000 | 0.799% | 10.31% | 17.28% | | |
| | Nov | 2,143 | 284,000 | 0.755% | 4.17% | -5.72% | | |
| | Dec | 2,135 | 314,000 | 0.680% | -6.14% | -10.42% | | |
| Model Inputs | | January '20 Rate 0.624% | | 5-year Annual Growth Rate 2.97% | | 5-year Volatility 68.61% | 5-year Volatility of Volatility 11.73% | |

## 7.2. Distribution of Crash Rates: 2009-2020

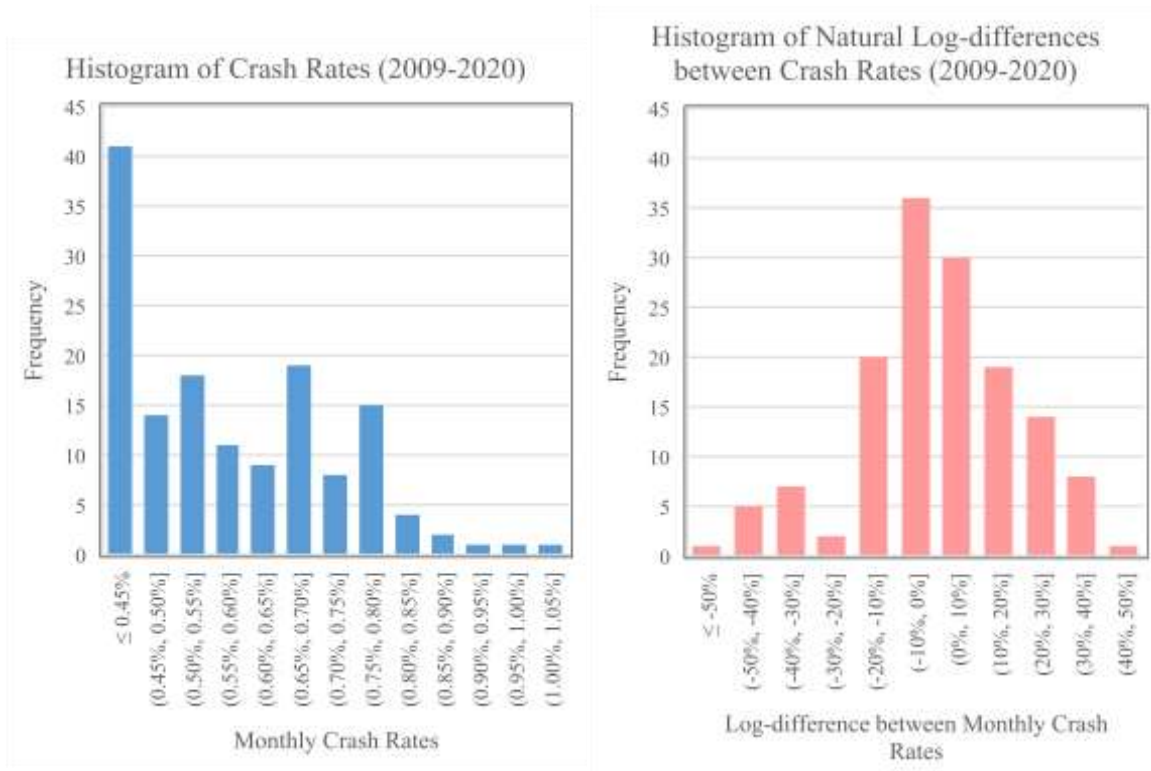

**Figure A1 Distribution of (left) crash rates and (right) natural log-differences between successive crash rates between 2009 and 2020. Crash rates follow a lognormal distribution, while the log-differences between rates follow a normal distribution.**